\documentclass[conference,twocolumn]{IEEEtran}
\IEEEoverridecommandlockouts
\usepackage{cite}
\usepackage{amsmath,amssymb,amsfonts}
\usepackage{algorithm}
\usepackage{algorithmic}
\usepackage{graphicx}
\usepackage{textcomp}
\usepackage{xcolor}
\usepackage{amsthm}
\usepackage{dsfont}
\usepackage{setspace}
\usepackage[T1]{fontenc}
\usepackage{aecompl}
\usepackage{url}

\def\BibTeX{{\rm B\kern-.05em{\sc i\kern-.025em b}\kern-.08em
    T\kern-.1667em\lower.7ex\hbox{E}\kern-.125emX}}
    
\newtheorem{theorem}{Theorem}
\newtheorem{lemma}{Lemma}

\begin{document}

\title{Optimal Update in Energy Harvesting Aided Terahertz Communications with Random Blocking\\
% {\footnotesize \textsuperscript{*}Note: Sub-titles are not captured in Xplore and
% should not be used}
%\thanks{%None}
% The appendix of this paper can be obtained through this link: {https://cloud.tsinghua.edu.cn/f/58ca1c5cd0404dffb639/} }
}

\author{
	\IEEEauthorblockN{Lixin Wang$^1$, Fuzhou Peng$^2$, Xiang Chen$^2$, \textit{Member, IEEE}, Shidong Zhou$^1$, \textit{Member, IEEE}}
	\IEEEauthorblockA{$^{1}$Department of Electronic Engineering, Tsinghua University, Beijing, Beijing 100084, China}
	\IEEEauthorblockA{$^{2}$School of Electronics and Information Technology, Sun Yat-sen University, Guangzhou, Guangdong 510006, China}
	\IEEEauthorblockA{wanglx19@mails.tsinghua.edu.cn, pengfzh@mail2.sysu.end.cn, chenxiang@mail.sysu.end.cn, zhousd@tsinghua.end.cn}
}
	
% \author{\IEEEauthorblockN{1\textsuperscript{st} Given Name Surname}
% \IEEEauthorblockA{\textit{dept. name of organization (of Aff.)} \\
% \textit{name of organization (of Aff.)}\\
% City, Country \\
% email address or ORCID}
% \and
% \IEEEauthorblockN{2\textsuperscript{nd} Given Name Surname}
% \IEEEauthorblockA{\textit{dept. name of organization (of Aff.)} \\
% \textit{name of organization (of Aff.)}\\
% City, Country \\
% email address or ORCID}
% \and
% \IEEEauthorblockN{3\textsuperscript{rd} Given Name Surname}
% \IEEEauthorblockA{\textit{dept. name of organization (of Aff.)} \\
% \textit{name of organization (of Aff.)}\\
% City, Country \\
% email address or ORCID}
% \and
% \IEEEauthorblockN{4\textsuperscript{th} Given Name Surname}
% \IEEEauthorblockA{\textit{dept. name of organization (of Aff.)} \\
% \textit{name of organization (of Aff.)}\\
% City, Country \\
% email address or ORCID}
% \and
% \IEEEauthorblockN{5\textsuperscript{th} Given Name Surname}
% \IEEEauthorblockA{\textit{dept. name of organization (of Aff.)} \\
% \textit{name of organization (of Aff.)}\\
% City, Country \\
% email address or ORCID}
% \and
% \IEEEauthorblockN{6\textsuperscript{th} Given Name Surname}
% \IEEEauthorblockA{\textit{dept. name of organization (of Aff.)} \\
% \textit{name of organization (of Aff.)}\\
% City, Country \\
% email address or ORCID}
% }

\maketitle

\begin{abstract}
In this paper, we consider an information update system where wireless sensor sends timely updates to the destination over a random blocking terahertz channel with the supply of harvested energy and reliable energy backup. The paper aims to find the optimal information updating policy that minimize the time-average weighted sum of the Age of information(AoI) and the reliable energy costs by formulating an infinite state Markov decision process(MDP). With the derivation of the monotonicity of value function on each component, the optimal information updating policy is proved to have a threshold structure. Based on this special structure, an algorithm for efficiently computing the optimal policy is proposed. Numerical results show that the optimal updating policy proposed outperforms baseline policies.
% Age of Information(AoI), the  time  elapsed  since  the  destination  received  the  latest information update, is adopted to measure the timeliness of the system. 
\end{abstract}


\section{Introduction}
\label{section1}
%  As an important part of the Internet of Things(IoT), information update systems based on wireless sensors have attracted much attention. Through this update system, the sensor will send updated information to the destination, so that the destination can perceive the changes in a wider range of environment in time and make a reasonable decision. Therefore, the timeliness of the update is very important. 
 Timely information updates from wireless sensors to the destination are critical in real-time monitoring and control systems.
In order to describe the timeliness of information update, the metric Age of Information(AoI) is proposed\cite{kaul2012real}. Different from general performance metrics such as delay and throughput, AoI refers to the time elapsed since the destination received the latest information. A lower AoI usually reflects the more timely information is updated, which is also expected. However, due to the limited energy of wireless sensors and the uncertainty of the transmission channel, frequent information update is not necessarily the optimal information updating policy. Therefore, information updating policies under energy-constrained and random channel conditions have been widely studied\cite{sun2017update,yun2018optimal,peng2021channel}.
 
Meanwhile, energy harvesting, as a promising technology, is widely used in autonomous wireless sensor networks\cite{ma2019sensing}. Energy harvesting can continuously replenish energy for the sensor by extracting energy from solar power, ambient RF and thermal energy. Many works are based on the setting of energy harvesting to design information updating policy\cite{wu2017optimal,bacinoglu2018achieving,arafa2019age}. In \cite{wu2017optimal}, the author discussed the impact of the capacity of the battery used to store harvesting energy on the optimal updating policy. 
%When the battery capacity tends to infinity, a best effort uniform status update policy should be adopted. 
%Especially 
When the battery size is one, the optimal policy is proved to have a threshold structure. Then in \cite{bacinoglu2018achieving}, this result is generalized to any integer battery capacity. Further, in \cite{arafa2019age}, the random battery recharge (RBR) model and incremental battery recharge(IBR) model are considered to minimize AoI for data transmission of energy harvesting sensors. However, these studies did not take into account the limitations of energy harvesting. For example, in some systems that require periodic information updates, wireless sensors may not have enough energy to send updates due to the uncertainty of energy arrival. Wireless sensors that rely solely on energy harvesting to update information are unreliable.  
% Due to the uncertainty of harvesting energy, sensors that rely solely on harvesting energy for information update are unreliable. For example, the sensor may not have enough energy to update for a long period when there is no energy to arrive.

Therefore, it is necessary to consider a mixed energy supply mode in which reliable energy backup and harvested energy coexist\cite{jackson2019capacity,instruments2019bq25505,wu2020optimal,wu2020delay,draskovic2021optimal}. This kind of design is not only researched by academia\cite{jackson2019capacity}, but also actively promoted by industry\cite{instruments2019bq25505}. The additional reliable energy backup can enable the entire EH-aided wireless sensor to operate without interruption and increase the speed of cold start from empty energy storage. Nevertheless, at the same time, this also brings about the problem of energy management. Since harvested energy is free to use while the use of reliable energy requires a price, the wireless sensor should make full use of the harvested energy and minimize the backup energy consumption\cite{wu2020optimal,wu2020delay,draskovic2021optimal}. However, in the information update system, keeping the data fresh while minimizing the cost of using reliable energy is still an open problem.

In order to solve this problem, we consider a point-to-point information update system where the sensor can use both harvested energy and reliable energy to send information updates to the destination through a wireless channel. 
% In order to solve this problem, we consider a point-to-point information updating system where the wireless sensor can use both harvested energy and reliable energy to send information updates to the destination through a terahertz channel. 
Since the amount of data contained in an update may be very large, it is necessary to consider terahertz communication, which can provide greater bandwidth and higher transmission rates. 
The terahertz communication has a probability of being blocked by moving objects, so it can be modeled as a random blocking channel\cite{wu2020interference}. This paper will minimize the long-term average weighted sum of the AoI and the paid energy costs to find the optimal information updating policy. The key contributions of this paper are as follows:

\begin{itemize}
\item Modeling the above problem as an infinite state Markov decision process(MDP), and by showing the monotonicity of the value function on each component, we prove the threshold structure of the optimal policy.
\item 
% A modified value iterative algorithm is proposed, which converts the solution of the optimal policy in an infinite state space into a solution in a finite state set, which greatly reduces the complexity of solving the optimal policy. 
An efficient algorithm for solving the optimal policy is proposed based on the known threshold structure. 
% , which converts the solution of the optimal policy in the infinite state space into a finite state set. 
% The simulation results verify the correctness of the previous mathematical derivation, and show that the performance of the optimal policy is better than the zero-wait policy and the periodic policy. 
The simulation results verify the threshold  structure and show the influence of system parameters on the performance of the optimal policy. The optimal policy always performs better than the zero-wait policy and the periodic policy.
% show  the  threshold  structure  of  optimal  policy  and  comparethe  performance  affected  by  the  system  parameters
% Moreover, simulation results verify the correctness of the algorithm and the performance of the optimal policy.
% Moreover, the correctness of proposed algorithm is verified by subsequent simulations.
\end{itemize}

% The rest of the paper is organized as follows: Section \ref{section2} is about the introduction of the system model and problem formulation. Section \ref{section3} is the structure analysis of the optimal policy. Numerical simulation is presented in Section \ref{section4}. Finally, we draw the conclusion in Section \ref{section5}.

\section{SYSTEM MODEL and Problem Formulation}
\label{section2}
\subsection{System Model Overview}

\begin{figure}[tbp]
\centerline{\includegraphics[width=0.5\textwidth]{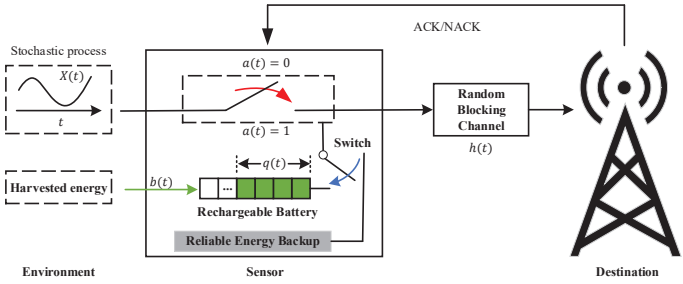}}
\caption{System model.}
\label{system}
\end{figure}

In this paper, we consider an information update system consisting of an EH-aided wireless sensor and a destination, as shown in Fig.~\ref{system}. With the supply of free harvest energy in the rechargeable battery and paid reliable energy backup, the sensor generates and transmits real-time environmental information updates to the destination over a random blocking channel. And there is a noise-free acknowledgement feedback channel from the destination to the sensor.

Without loss of generality, time is slotted with equal length and normalized to unity. In each time slot, the sensor decides whether to generate and transmit an update to the destination or stay idle. The decision action, denote by $a[t]$, takes value from action set $\mathcal{A}=\left\{0,1\right\}$. In time slot $t$, $a[t] =1$ means that the sensor decide to generate and transmit an update to the destination over the wireless blocking channel while $a[t]=0$ means the sensor is idle. The destination will feed back an ACK to the sensor when it has successfully received an update and a NACK otherwise. Note that according to our assumptions, the above processes can be completed in one time slot.

% In each time slot, one generated data packet arrives at the transmission buffer. If it is not scheduled to transmit in this slot, then it will be cleared from the buffer, because the packet in the next slot is fresher and can help the receiver obtain more timely environmental information. If the sensor transmits an update packet in this slot, an ACK will be fed back when base station successfully receives the data and a NACK otherwise.

%sampler of the sensor samples a stochastic process $X(t)$, generates and sends the sampled data to the transmission buffer of sensor, where it waits for transmission. Both the free harvested energy and paid reserved energy can provide the energy consumed by the packet transmission. Scheduler aims to minimize destination-AoI and paid power by scheduling the data transmission. The scheduling action, denote by $a[t]$, takes value from action set $\mathcal{A}=\left\{0,1\right\}$. In slot $t$, $a[t] =1$ means that sensor decide to transmit a data packet to the destination over the wireless block channel while $a[t]=0$ means its transmission unit is idle.

\subsection{Age of Information}
% of Information (AoI), is a freshness metric which is defined as the time passed since the generation time of the latest update packet available at the destination in this paper.
Age of Information (AoI), a freshness indicator, is defined as the elapsed time since the destination received the latest update in this paper. 
%Let $G_i$ denotes the generation time of the $i$th successfully received data packet,
Let $U[t]$ is the time slot of the latest update received by the destination before time slot $t$, $\Delta[t]$ denotes the AoI of destination in time slot $t$. Then, the AoI is given by
% \begin{equation}
% {\Delta[t]} = t - \max \{ {G_i}:{G_i} < t\}.
% \label{AoI}
% \end{equation}
\begin{equation}
{\Delta[t]} = t - U[t].
\label{AoI}
\end{equation}

In particular, the AoI will decreases to one if a new update is successfully received. Otherwise it will increase by one. To summarize, the evolution of AoI can be expressed as follows:

\begin{equation}
\label{equ_evolving_AoI}
\Delta[t+1]=
\begin{cases}
1, &\text{ successful transmission},\\
\Delta[t]+1, &\text{ otherwise}.
\end{cases}
\end{equation}

\subsection{Description of Energy Supply}

%The sensor can harvest energy from the environment in each slot. The collected energy is quantified as the number of energy packets, which can be used for data transmission free of charge. 
The EH-aided wireless sensor can send updates with the energy supply from energy harvesting and reliable energy backup. The harvested energy is quantified as energy packets. Denote $b[t]$ as the number of energy packets at the beginning of time slot $t$. It is assumed that the arrival process of energy packets is a Bernoulli Process with parameter $\lambda$. The distribution of $b[t]$ is as follows:
\begin{equation}
\label{b_distribution}
\begin{cases}
\Pr\left\{b[t]=k_1\right\}=\lambda,\\
\Pr\left\{b[t]=0\right\}=1-\lambda,
\end{cases}
\end{equation}where $\lambda \in (0,1]$ and $k_1 \in \mathbb Z^+$. Assume that all the harvested energy is stored in a rechargeable battery with an initial capacity of 0 in the sensor. The maximum storage capacity of the battery is $B$ (we assume that $B>1$). When the stored energy reaches $B$, the battery can not store the coming energy unless the sensor consumes battery energy to generate and send an update  in this time slot. Let $q[t]$ denotes the battery state, which means the stored energy which can be provided for data generation and transmission in time slot $t$. Then $q[t]$ takes value from the set $\mathcal{B}=\left\{0,1,...,B\right\}$. Generally, the sensor will give priority to using free harvesting energy for possible information updates and each information update consumes one harvested energy packet. Due to the randomness of energy harvesting in the changing environment, the battery energy may sometimes be insufficient to support update. Therefore, when the battery is empty, the sensor will automatically switch to the system's reliable energy backup. It is worth noting that the system can still harvest energy when battery state $q[t]=0$. The evolution of battery state between time slot $t-1$ and $t$ can be summarized as follows:
\begin{equation}
\label{q_evolution}
q[t] = \max \left\{ \min \{ q[t - 1] + b[t] - a[t]u(q[t-1]),B\} ,0 \right\},
\end{equation}
where $u(\cdot)$ is unit step function, which is defined as
\begin{equation}
\label{indicator}
u(x)=
\begin{cases}
1,&\text{if $x>0$},\\
0,&\text{otherwise}.
\end{cases}
\end{equation}

\subsection{Channel Model}
%As is known the communication spectrum used by terahertz communication is too high, it is more suitable for LOS communication. In the case of obstruction by objects, communication may be interrupted. For this characteristic, we can use random blocking channels to model. 

A random blocking channel model is used to characterize the terahertz channel. Let $h[t]$ denote the state of channel in time slot $t$. The channel has two states: ‘block’ and ‘unblock’, the corresponding probabilities are $p$ and $1-p$, respectively. Note that $p  \in (0,1)$. So the probability distribution of blocked channels is given by:
\begin{equation}
\label{ht}
\begin{cases}
{\rm{Pr}}\left\{h[t]={\text{‘block’}}\right\}=p,\\
{\rm{Pr}}\left\{h[t]={\text{‘unblock’}}\right\}=1-p.
\end{cases}
\end{equation}
It is assumed that the channel blocking probability distributions are \textit{i.i.d.} in different time slots.

\subsection{Problem Formulation}
%Based on the known AoI and battery status information, the sensor makes a real-time decision to decide whether to perform data transmission in each time slot. This paper aims to find an online optimal policy that minimizes the weighted sum of system AoI and paid energy consumption, or we say average cost in the long term. 
This paper aims to find the optimal information updating policy that achieves the minimum of the time-average weighted sum of the AoI and the paid reliable energy costs. Let $\Pi$ denotes the set of the stationary and deterministic policies. For any $\pi\in \Pi$, it can be represented by a sequence of actions, i.e., $\pi = ({a[0]},{a[1]},{a[2]},...{a[t]},...)$. Suppose that under paid reliable energy supply, the cost of generating and transmitting an information update is a non-negative value $C_r$,
% Assuming that each unit of paid energy consumption is a non-negative value $C_r$ 
then we formulate our problem as follows:
\begin{equation}
\label{problem}
\mathop {\min }\limits_{\pi  \in \Pi } \mathop {\lim }\limits_{T \to \infty } \sup \frac{1}{T}{\mathbb E}\left\{\sum\limits_{t = 0}^{T-1}  [{\Delta [t]} + \omega{C_r}a[t]\mathds{1}(q[t])]\right\},
\end{equation}
where $\omega$ is the positive weighting factor and $\mathds{1}(\cdot)$ is indicator function defined as follows:

\begin{equation}
\label{indicator_1}
\mathds{1}(x)=
\begin{cases}
1,&\text{if $x=0$},\\
0,&\text{otherwise}.
\end{cases}
\end{equation}
It appears here because senor may use paid reliable energy and generate corresponding cost only when the battery state $q[t]$ is 0.

%By solving this problem, we can get the optimal policy we need.However, it is almost impossible to solve this problem directly. In the next section, we will model the problem as a Markov decision process in an infinite state space, and analyze the optimal solution structure and obtain the optimal policy.

\section{Optimal policy analysis}
\label{section3}
In this section, we aim to solve the problem (\ref{problem}) and obtain the optimal policy. The original problem is first reformulated as a time-average cost MDP with infinite state space.
%and its state transition probability model is established according to the description of the previous system model. %Then by verifying that it satisfies the theorem in [16], this problem has a stationary and deterministic optimal policy that minimizes the average cost.%
By analysing the properties of the value function, we prove that the optimal policy is of a threshold structure related to AoI with a given battery state. Moreover, this paper also proposes a modified value iteration algorithm (VIA) based on the known threshold structure to reduce the computational complexity of finding the optimal policy.

\subsection{Markov Decision Process Formulation}
Markov decision process is typically used to model and analyze model-based sequential decision problems with per-stage cost. According to the system description mentioned on the previous section, the MDP is formulated as follows:

\begin{itemize}
\item \textbf{State Space}. The state of a sensor $\textbf{x}[t]$ in slot $t$ is a couple of the current destination-AoI and the battery state, i.e., $(\Delta[t], q[t])$. The state space $\mathcal{S}= \mathbb Z^+ \times \mathcal{B}$ is thus inﬁnite countable.
\item \textbf{Action Space}. The sensor's action $a[t]$ in time slot $t$ only takes value from the action set $\mathcal{A}=\left\{0,1\right\}$. 
\item \textbf{Transition Probability}. Denote $\Pr ({\textbf{x}}[t + 1]|{\textbf{x}}[t],a[t])$ as the transition probability that current state $\textbf{x}[t]$ transits to next state $\textbf{x}[t+1]$ after taking action $a[t]$. When $k_1=1$, the transition probability is divided into two cases conditioned on different values of action.

\textbf{\emph{Case 1}}. $a[t]=0$,
\begin{equation}
\label{transition_case1_v2}
\begin{cases}
\Pr \{ (\Delta+1, q+1)|(\Delta, q),0\}=\lambda, &\text{ if }q < B, \\
\Pr \{(\Delta+1, B)|(\Delta, B),0\}= 1,         &\text{ if } q = B, \\
\Pr \{(\Delta+1, q)|(\Delta, q),0\}= 1-\lambda, &\text{ for all }q. \\
% \Pr ({\textbf{x}}[t\!+\!1]\!=\!(\Delta\!+\!1, q)|{\textbf{x}}[t]\!=\!(\Delta, q),a[t]\!=\!0)\!=\!1\!-\!\lambda, \\
% \Pr ({\textbf{x}}[t\!+\!1]\!=\!(\Delta\!+\!1, q\!+\!1)|{\textbf{x}}[t]\!=\!(\Delta, q),a[t]\!=\!1)\!=\!p\lambda, \\
% \Pr ({\textbf{x}}[t\!+\!1]\!=\!(\Delta\!+\!1, q)|{\textbf{x}}[t]\!=\!(\Delta, q),a[t]\!=\!1)\!=\!p(1-\!\lambda), \\
% \Pr ({\textbf{x}}[t\!+\!1]\!=\!(1, q\!+\!1)|{\textbf{x}}[t]\!=\!(\Delta, q),a[t]\!=\!1)\!=\!(1\!-\!p)\lambda, \\
% \Pr ({\textbf{x}}[t\!+\!1]\!=\!(1, q)|{\textbf{x}}[t]\!=\!(\Delta, q),a[t]\!=\!1)\!=\!(1\!-\!p)(1\!-\!\lambda).
\end{cases}
\end{equation}
In this case, the evolution of AoI follows form equation \eqref{equ_evolving_AoI}. The evolution of the battery state follows equation \eqref{q_evolution}. It is worth noting that when the harvested energy is $B$, the arrival energy can not be stored in the rechargeable battery. 

\textbf{\emph{Case 2}}. $a[t]=1$,
\begin{equation}
\label{transition_case2_v2}
\begin{cases}
\Pr \{ (\Delta+1, q)|(\Delta, q),1\}=p\lambda,              &\text{ if }q>0, \\
\Pr \{ (1, q)|(\Delta, q),1\}=(1-p)\lambda,                 &\text{ if }q>0, \\
\Pr \{\Delta+1, q-1)|(\Delta, q),1\}=p(1-\lambda),           &\text{ if }q>0, \\
\Pr \{ (1, q-1)|(\Delta, q),1\}=(1-p)(1-\lambda),           &\text{ if }q >0, \\
\Pr \{ (\Delta+1, 1)|(\Delta, 0),1\}=p\lambda,              &\text{ if }q=0, \\
\Pr \{ (1, 1)|(\Delta, 0),1\}=(1-p)\lambda,                 &\text{ if }q=0, \\
\Pr \{(\Delta+1, 0)|(\Delta, 0),0\}= p(1-\lambda),          &\text{ if } q = 0, \\
\Pr \{(1, 0)|(\Delta, 0),0\}= (1-p)(1-\lambda),             &\text{ if } q = 0. \\
% \Pr ({\textbf{x}}[t\!+\!1]\!=\!(\Delta\!+\!1, q)|{\textbf{x}}[t]\!=\!(\Delta, q),a[t]\!=\!0)\!=\!1\!-\!\lambda, \\
% \Pr ({\textbf{x}}[t\!+\!1]\!=\!(\Delta\!+\!1, q\!+\!1)|{\textbf{x}}[t]\!=\!(\Delta, q),a[t]\!=\!1)\!=\!p\lambda, \\
% \Pr ({\textbf{x}}[t\!+\!1]\!=\!(\Delta\!+\!1, q)|{\textbf{x}}[t]\!=\!(\Delta, q),a[t]\!=\!1)\!=\!p(1-\!\lambda), \\
% \Pr ({\textbf{x}}[t\!+\!1]\!=\!(1, q\!+\!1)|{\textbf{x}}[t]\!=\!(\Delta, q),a[t]\!=\!1)\!=\!(1\!-\!p)\lambda, \\
% \Pr ({\textbf{x}}[t\!+\!1]\!=\!(1, q)|{\textbf{x}}[t]\!=\!(\Delta, q),a[t]\!=\!1)\!=\!(1\!-\!p)(1\!-\!\lambda).
\end{cases}
\end{equation}

In this case, the evolution of AoI still follows from equation \eqref{equ_evolving_AoI}. The evolution of the battery state should be discussed by two situations, i.e., $B\geq q>0$ and $q=0$. In the first situation, the battery state follows \eqref{q_evolution}. While in another situation, if there harvests a unit energy, the battery state increases by one due to that the sensor uses the paid reserved energy in this slot. Otherwise, the battery state keeps zero. When $k_1>1$, we can get the transition probability model through the same steps. Note that in the rest of the paper we will focus on this transition probability model where $k_1=1$.

\item \textbf{One-step Cost}. For the current state $\textbf{x}=(\Delta, q)$, the one-step cost $C(\textbf{x},a)$ of taking action $a$ is expressed by
%includes the AoI growth and the possible paid energy cost. The specific expression is as follows:
 
\begin{equation}    
    \label{onestepcost}
    C(\textbf{x},a) = \Delta  + \omega {C_r}a\mathds{1}(q).
\end{equation}
\end{itemize}

After the above modeling, the original problem \eqref{problem} is transformed into obtaining the optimal policy for the MDP to minimize the average cost in an infinite horizon:
\begin{equation}
\label{trans_problem}
 \mathop {\lim }\limits_{T \to \infty } \sup \frac{1}{T}{\mathbb E_\pi}\left\{ \sum\limits_{t = 0}^{T-1} C(\textbf{x}[t],a[t])\right\}.
\end{equation}

According to \cite{altman1999constrained} , a stationary deterministic policy to minimize the above unconstrained MDP with inﬁnite countable state and action space exists under certain verifiable conditions. The next section, the structure properties of optimal policy is investigated.
%The next section gives a rigorous theoretical analysis and proof of the structure of the optimal policy for this problem.
\subsection{Structure Analysis of Optimal Policy}
In this section, some preliminary lemmas are established to reveal the properties of value function. Based on these, it is proved that the optimal policy is of a threshold structure. Therefore, an efficient algorithm, naming modified value iteration algorithm, for obtaining the optimal policy based on the threshold structure will be presented.

%first we will give some lemmas to prove the threshold structure of the optimal policy. Then an efficient algorithm for obtaining the optimal policy based on the threshold structure will be presented.
According to \cite{sennott1989average}, there exits a value function $V(\textbf{x})$ which satisﬁes the following Bellman equation for the infinite horizon average cost MDP:
\begin{equation}
\lambda + V(\mathbf{x}) = \min_{a \in \mathcal{A}} \left\{ C(\mathbf{x},a) + \sum_{\mathbf{x}^\prime \in \mathcal{S}} \Pr (\mathbf{x}^\prime|\mathbf{x},a)V(\mathbf{x}^\prime) \right\},
\label{bel_equation}
\end{equation}
where $\lambda$ is the average cost by following the optimal policy. Denote $Q(\mathbf{x},a)$ as the state-action value function which means the value of taking action $a$ in state $\mathbf{x}$. We have:
\begin{equation}
    Q(\mathbf{x},a)=C(\mathbf{x},a) + \sum_{\mathbf{x}^\prime \in \mathcal{S}} \Pr (\mathbf{x}^\prime|\mathbf{x},a)V(\mathbf{x}^\prime).
\label{Qfunction}
\end{equation}

So the optimal policy in state $\mathbf{x}$ can be expressed as follows:
\begin{equation}
    {\pi ^\star}(\mathbf{x}) = \arg \mathop {\min }\limits_{a \in \mathcal{A}} Q(\mathbf{x},a).
\label{piandQequation}
\end{equation}

Next, we first prove the monotonicity of the value function on different dimensions, which is summarized in the following lemma.

\begin{lemma}
    For a fixed channel blocking probability $p$, given the battery state $q$ and for any $1\leq \Delta_1\leq\Delta_2$, we have 
    %the value function monotonically non-decreasing with the increase of AoI; for a fixed AoI, the value function monotonically non-increasing with the increase of $q$. That is, for any $\Delta_1\le\Delta_2$ and fixed $q \in \mathcal{B}$, we have:
    \begin{equation}
    \label{lemma1_part1}
        V(\Delta_1,q)\le V(\Delta_2,q),
    \end{equation}
    %For any $q \in \left\{0,1,...,B-1\right\}$ and $\Delta \in \mathbb Z^+ $,we have:
    and, given AoI $\Delta\geq 1$,
    \begin{equation}
        \label{lemma1_part2}
        V(\Delta,q)\ge V(\Delta,q+1)
    \end{equation}
    holds for any $q\in \left\{0,1,...,B-1\right\}$.
    \label{lemma1}
\end{lemma}

\begin{IEEEproof}
See Appendix \ref{app_proof_lemma_monitonic} in Supplementary Material \cite{tsinghua.edu}.
\end{IEEEproof}

Based on Lemma \ref{lemma1}, we then establish the incremental property of the value function, which is shown in the following lemma.

%as the state changes, which is shown in the following lemma. 
\begin{lemma}
    For a fixed channel blocking probability $p$, for any $\Delta_1 \le \Delta_2$ and given $q \in \mathcal{B}$, we have:
    \begin{equation}
    \label{lemm2_formula1}
        V(\Delta_2,q)-V(\Delta_1,q)\ge \Delta_2-\Delta_1.
    \end{equation}
    And, for any $q \in \left\{0,1,...,B-1\right\}$ and $\Delta \in \mathbb Z^+ $, we have:
    \begin{equation}
    \label{lemm2_formula2}
        V(\Delta+1,q+1)-V(\Delta,q+1)\ge p[V(\Delta+1,q)-V(\Delta,q)].
    \end{equation}
    \label{lemma2}    
\end{lemma}
\begin{IEEEproof}
See Appendix \ref{app_proof_lemma_creasement} in Supplementary Material \cite{tsinghua.edu}. 
\end{IEEEproof}

With Lemma \ref{lemma1} and Lemma \ref{lemma2}, we directly provide our main result in the following Theorem.

%prove the threshold structure of the optimal policy of the MDP, i.e., Theorem \ref{theorem1}.
     
\begin{theorem}
    Assuming that the channel blocking probability $p$ is fixed. For given battery state $q$, there exists a threshold $\Delta_q$ , such that when $\Delta\ < \Delta_q$, the optimal action $\pi^\star (\Delta,q)=0$, i.e., the sensor keeps idle; when $\Delta \ge \Delta_q$, the optimal action $\pi^\star(\Delta,q)=1$, i.e., the sensor chooses to generate and transmit a new update.
\label{theorem1}
\end{theorem}
\begin{IEEEproof}
The optimal policy is of a threshold structure if $Q(\mathbf{x},a)$ has a sub-modular structure, that is,
\begin{equation}
Q(\Delta,q,0)- Q(\Delta,q,1) \leq  Q(\Delta+1,q,0)- Q(\Delta+1,q,1).
\end{equation}

We will divided the whole proof by the following three cases:
    
    %Since the optimal policy is obtained by $    {\pi ^\star}(\mathbf{x}) = \arg \mathop {\min }\limits_{a \in \mathcal{A}} Q(\mathbf{x},a)$,we can compare the value of different actions in every state. We will discuss it in the following three cases:
    
    \textbf{Case 1}. When $q=0$, for any $\Delta \in \mathbb Z^+$ we have:
    \begin{align}
        &Q(\Delta,q,0)-Q(\Delta,q,1)\nonumber\\
        =&\Delta+ \lambda V(\Delta+1,q+1)+(1-\lambda)V(\Delta+1,q)\nonumber\\
        &-\Delta-\omega{C_r}-p\lambda V(\Delta+1,q+1)+p(1- \lambda)V(\Delta+1,q)\nonumber\\
    &-(1-p)\lambda V(1,q+1)-(1-p)(1-\lambda) V(1,q)\nonumber\\
    =&(1-p)\lambda(V(\Delta+1,q+1)-V(1,q+1))\nonumber\\
    &+(1-p)(1-\lambda)(V(\Delta+1,q)-V(1,q))-\omega{C_r}.
    \end{align}
    %&\ge (1-p)\Delta-C_r,
    %where the last inequality is due to the incremental property revealed by \eqref{lemm2_formula1} in Lemma \ref{lemma2}. Since $q \in (0,1)$ and $Ce>0$, $(1-p)\Delta-C_r$ increases linearly with $\Delta$.Then there exists an constant value $\Delta_0$ satisfies $(1-p)\Delta_0-C_r=0$. Therefore, if $\Delta\ge\Delta_0$, $Q(\Delta,q,0)-Q(\Delta,q,1)\ge(1-p)\Delta_0-C_r=0$, the optimal policy is to transmit data, otherwise, keep idle.
    
Therefore, we have
\begin{align}
&Q(\Delta+1,q,0)-Q(\Delta+1,q,1) - [Q(\Delta,q,0)-Q(\Delta,q,1)]\nonumber\\
=&(1-p)\lambda(V(\Delta+2,q+1)-V(\Delta+1,q+1)) \nonumber\\
&+(1-p)(1-\lambda)(V(\Delta+2,q)-V(\Delta,q))\nonumber\\
\geq& 0,
\end{align}
where the last inequality is due to the monotonicity property revealed by \eqref{lemma1_part1} in Lemma \ref{lemma1}. This completes the proof of this case.

    \textbf{Case 2}. When $q \in \left\{1,...,B-1\right\}$,for any $\Delta \in \mathbb Z^+$ we have:
    \begin{align}
        &Q(\Delta+1,q,0)-Q(\Delta+1,q,1)-[Q(\Delta,q,0)-Q(\Delta,q,1)]\nonumber\\
        =&Q(\Delta+1,q,0)-Q(\Delta,q,0)-[Q(\Delta+1,q,1)-Q(\Delta,q,1)]\nonumber\\
        =&\lambda[V(\Delta+2,q+1)-V(\Delta+1,q+1)]\nonumber\\
        &-p\lambda[V(\Delta+2,q)-V(\Delta+1,q)]\nonumber\\
        &+(1-\lambda)[V(\Delta+2,q)-V(\Delta+1,q)]\nonumber\\
        &-p(1-\lambda)[V(\Delta+2,q-1)-V(\Delta+1,q-1)]\nonumber\\
        \ge& 0,
        \label{submodular}
    \end{align}    
    where the last inequality is due to the incremental property revealed by \eqref{lemm2_formula2} in Lemma \ref{lemma2}. This completes the proof of this case.
    
    %The formula \eqref{submodular} indicates that if $\Delta_q$ is the threshold, so is $\Delta_q+1$. Since the optimal policy is not to keep idle all the time, there must exists a thresold $\Delta_q$ with respect to $q$.
    
    \textbf{Case 3}. When $q=B$,for any $\Delta \in \mathbb Z^+$ we have:
    \begin{align}
        &Q(\Delta+1,q,0)-Q(\Delta+1,q,1)-[Q(\Delta,q,0)-Q(\Delta,q,1)]\nonumber\\
        =&Q(\Delta+1,q,0)-Q(\Delta,q,0)-[Q(\Delta+1,q,1)-Q(\Delta,q,1)]\nonumber\\
        =&(1-\lambda)[V(\Delta+2,q)-V(\Delta+1,q)]\nonumber\\
        &-p(1-\lambda)[V(\Delta+2,q-1)-V(\Delta+1,q-1)]\nonumber\\
        \ge& 0,
        \label{B_submodular}
    \end{align}     
    where the last inequality is also due to the incremental property revealed by \eqref{lemm2_formula2} in Lemma \ref{lemma2}.
    
    %So the analysis here is the same as case 2 when $q \in \left\{1,...,B-1\right\}$. That is there exists a $\Delta_B$ as a  AoI-threshold for battery state $B$.
    
Therefore, we have completed the whole proof.
\end{IEEEproof}
Theorem \ref{theorem1} shows that if the optimal action in a certain state is to generate and transmit update, then in the state with the same battery state and larger AoI, the optimal action must be the same. Based on this unique threshold structure, we propose a modified value iteration algorithm that efficiently reduce the computational complexity of solving the optimal policy. See Algorithm \ref{mvia} for details.

\begin{algorithm}[tb]
    \caption{Modified Value Iteration Algorithm}
    \label{mvia}
    \begin{algorithmic}[1]
        \REQUIRE ~~\\ 
        Iteration number $k$ and iteration threshold: $\epsilon$.
        % Initial value of value function $V_0(\mathbf{x})$: $m_{\mathbf{x}}$;\\
        % Value function $V_1(\mathbf{x})$: $m_{\mathbf{x}}+2\epsilon$;\\    
        % Initial value of policy $\pi(\mathbf{x})$: 0;\\
        % Initial value of iteration number $k$: 0;\\
        % Iteration threshold: $\epsilon$.
        \ENSURE ~~\\ 
        Optimal policy $\pi^\star(\mathbf{x})$ for all state $\mathbf{x}$.
        \STATE \textbf{Initialization: }$V_0(\mathbf{x})= m_{\mathbf{x}}.$
        %\STATE  
        \WHILE{$k > 0$}
            \STATE $Q_k(\mathbf{x},a)\leftarrow C(\mathbf{x},a) +\underset{\mathbf{x}^\prime \in \mathcal{S}}{\sum}   \Pr (\mathbf{x}^\prime|\mathbf{x},a)V_k(\mathbf{x}^\prime)$ 
            \STATE ${V_{k + 1}}(\mathbf{x}) \leftarrow  \mathop {\min }\limits_{a \in \mathcal{A}} Q_k(\mathbf{x},a)$
            \IF{$\|V_{k+1}(\mathbf{x})-V_{k}(\mathbf{x})\|\leq \epsilon$}
            \STATE \textbf{break;}
            \ELSE
            \STATE $k\leftarrow k+1$
            \ENDIF
        \ENDWHILE
        \FOR{$\mathbf{x}=(\Delta,q) \in \mathcal{S}$}
            \IF{$\pi(\Delta-1,q)=1$}
            \STATE $\pi(\mathbf{x})\leftarrow 1$,
            \ELSE
            \STATE ${\pi}(\mathbf{x}) \leftarrow \arg \mathop {\min }\limits_{a \in \mathcal{A}} Q(\mathbf{x},a)$               
            \ENDIF
        \ENDFOR
        \STATE $\pi^\star(\mathbf{x})\leftarrow\pi(\mathbf{x})$
    \end{algorithmic}
\end{algorithm}

\section{Numerical Result}
\label{section4}
In this section, the simulation results are presented to show the threshold structure of optimal policy and compare the performance affected by the system parameters. In our simulation, we assume that cost of reliable energy for one update is $C_r=2$ and the maximum battery capacity $B=20$. 

%of optimal scheduling policy through numerical simulation, which also verifies our theoretical derivation. In our simulation, we assume energy price $C_r=2$.

%First, we verify the threshold structure of the optimal update policy. 

\begin{figure}[tbp]
    \centerline{\includegraphics[width=0.5\textwidth]{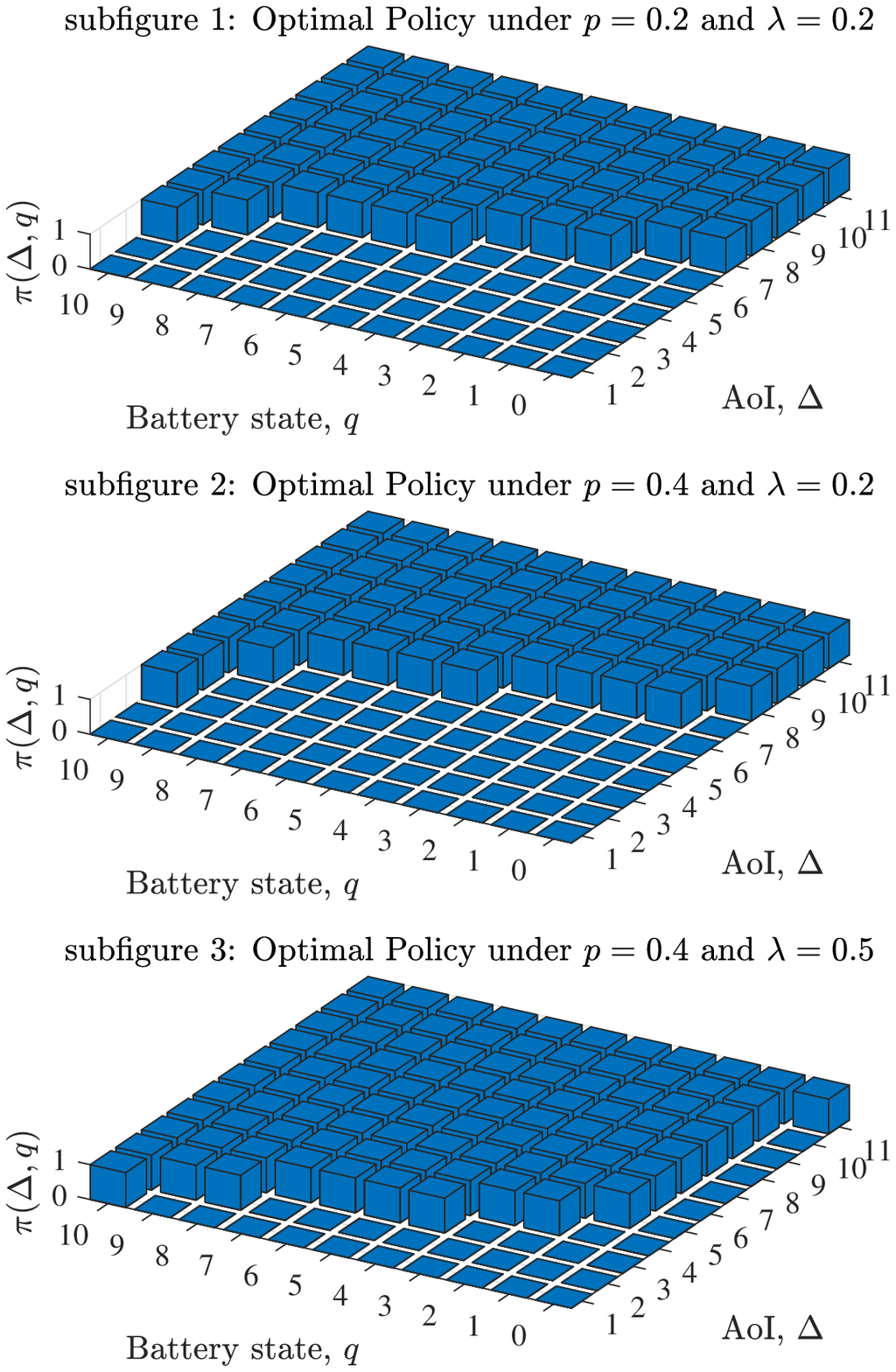}}
    \caption{Optimal policy conditioned on different parameters.}
    \label{fig:thresold_blue}
\end{figure}

Fig.~\ref{fig:thresold_blue} shows the optimal policy under different channel blocking probability and energy harvesting probability. Note that the weighting factor $\omega$ is set to be $10$. All the subfigures in Fig.~\ref{fig:thresold_blue} reflect the threshold structure. Comparing subfigure 1 and subfigure 2, we found that under the same energy harvesting probability, the greater the channel blocking probability $p$, the higher the threshold corresponding to each battery state. This is also in line with cognition, because as the channel uncertainty increases, the action of transmitting data may not necessarily bring about a reduction in AoI, but may move in the direction of consuming paid reliable energy. Comparing subfigure 2 and subfigure 3, under the same channel blocking probability, the greater the probability of energy harvesting, the threshold corresponding to most battery states will be reduced accordingly. An 'abnormal' phenomenon is that when $q=0$, the corresponding AoI threshold increases as the probability of energy harvesting increases. The reasonable explanation here is that the sensor is willing to pay the price of AoI growth to wait for the free harvested energy.
% to be used to generate and transmit update under the premise that only paid energy is available.  It can be found in the three subfigures that under the same other conditions, the AoI threshold decreases as the battery energy state increases, although this has not yet been theoretically proven.

\begin{figure}[tbp]
\centerline{\includegraphics[width=0.4\textwidth]{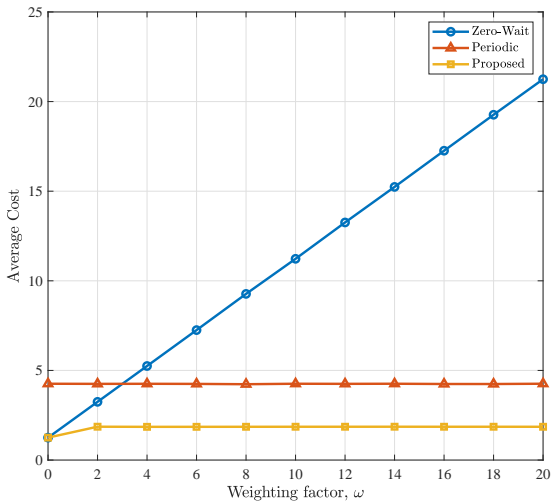}}
\caption{Performance comparison of the zero-wait policy, periodic policy and proposed policy versus the weighting factor $\omega$ with simulation conditions $p=0.2$, $\lambda=0.5$ and $B=20$.}
\label{costwithb_big}
\end{figure}

Then, we show the average cost performance of optimal policy in Fig.~\ref{costwithb_big} under different weighting factor $\omega$. Optimal policy is compared with zero-wait policy and the periodic policy (period = 5) under the same channel blocking probability $p=0.2$ and the energy harvesting probability $\lambda=0.5$ in this simulation. It can be found that under different weighting factor $\omega$, the optimal policy proposed in this paper can obtain the minimum long-term  average cost, compared with the other two policies. When $\omega$ tends to $0$, the zero-wait policy tends to the optimal policy. This is because when there is no need to consider the update cost brought by paid reliable energy, that is, when there is no energy consumption limit, the optimal policy is to update information in every time slot.

% Then, we show the optimal policy average cost performance in Fig.~\ref{costwithb_big} under different maximum battery size $B$. Optimal policy is compared with zero-wait policy and the periodic policy (period = 5) under the same channel block probability $p=0.2$ and the energy harvesting probability $\lambda=0.5$ in this simulation. It can be found that under different maximum battery capacities, the optimal policy proposed in this paper can obtain the minimum long-term  average cost, compared with the other two policies. As the maximum battery capacity increases, the performance gap between these algorithms has gradually stabilized. 
% It is worth noting that in the simulation, $B$ starts from 0 and 1, which is different from the previous assumption of $B>1$, but it can be regarded as a special case of the previous derivation, which can also obtain a consistent threshold structure and optimal policy.

\begin{figure}[tbp]
\centerline{\includegraphics[width=0.4\textwidth]{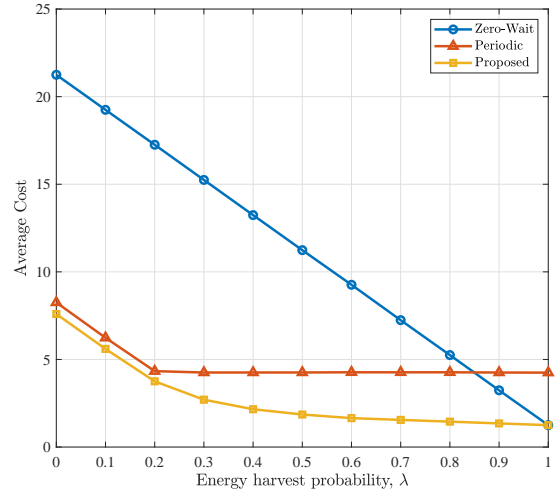}}
\caption{Comparison of the Zero-wait policy, periodic policy and proposed policy versus the energy harvesting probability with simulation conditions $p=0.2$, $\omega=10$ and $B=20$.}
\label{costwithlambda_big}
\end{figure}

In Fig.~\ref{costwithlambda_big}, we present the impact of different energy harvesting probabilities on different strategies. In this simulation, we set the channel blocking probability $p=0.2$ and weighting factor $\omega=10$. It can be found from the Fig.~\ref{costwithlambda_big} that for different energy harvesting probabilities, the proposed optimal update policy outperforms the zero-wait policy and the periodic policy(period = 5), that is, the long-term average cost is always smaller. The interesting point is that when the probability of energy harvesting tends to 1, that is, when energy arrives in each time slot, the performance of the zero-wait policy is close to the optimal policy, while  there is still a performance gap between the periodic policy and the optimal policy. This is predictable, because the optimal policy in that case must be to generate and transmit updates all the time without the need to use paid energy.
%with a steady supply of harvested energy. 
However the periodic policy can not make use of this information and wastes a lot of opportunities to update information without paying any cost.

% \begin{figure}[htbp]
%     \centerline{\includegraphics[width=0.5\textwidth]{fig/thresold_version2.eps}}
%     \caption{Caption}
%     \label{fig:thresold_blue}
% \end{figure}

\section{Conclusion}
\label{section5}
In this paper, we have studied the optimal updating policy for energy harvesting aided terahertz communication over random block channel. This scheduling problem has been transformed into an infinite state Markov decision process, and its goal is to minimize the long-term average weighted sum of the AoI and the energy consumption supplied by the paid energy. Some preliminary lemmas are first provided. Based on them, we prove that the optimal policy structure is of threshold type by exploiting the monotonicity of the value function. At the same time, an efficient policy search algorithm is proposed. Simulation results show that the threshold structure exists, and the thresholds are affected by the probability of energy harvesting and channel blocking. At the same time, it has been verified that the optimal policy is better than the zero-wait policy and the periodic policy.

%\section*{Acknowledgment}

% The preferred spelling of the word ``acknowledgment'' in America is without 
% an ``e'' after the ``g''. Avoid the stilted expression ``one of us (R. B. 
% G.) thanks $\ldots$''. Instead, try ``R. B. G. thanks$\ldots$''. Put sponsor 
% acknowledgments in the unnumbered footnote on the first page.

% \section*{References}

% Please number citations consecutively within brackets \cite{b1}. The 
% sentence punctuation follows the bracket \cite{b2}. Refer simply to the reference 
% number, as in \cite{b3}---do not use ``Ref. \cite{b3}'' or ``reference \cite{b3}'' except at 
% the beginning of a sentence: ``Reference \cite{b3} was the first $\ldots$''

% Number footnotes separately in superscripts. Place the actual footnote at 
% the bottom of the column in which it was cited. Do not put footnotes in the 
% abstract or reference list. Use letters for table footnotes.

% Unless there are six authors or more give all authors' names; do not use 
% ``et al.''. Papers that have not been published, even if they have been 
% submitted for publication, should be cited as ``unpublished'' \cite{b4}. Papers 
% that have been accepted for publication should be cited as ``in press'' \cite{b5}. 
% Capitalize only the first word in a paper title, except for proper nouns and 
% element symbols.

% For papers published in translation journals, please give the English 
% citation first, followed by the original foreign-language citation \cite{b6}.

\bibliographystyle{./IEEEtran}
\bibliography{./IEEEexample}

% \begin{thebibliography}{00}
% \bibitem{b1} G. Eason, B. Noble, and I. N. Sneddon, ``On certain integrals of Lipschitz-Hankel type involving products of Bessel functions,'' Phil. Trans. Roy. Soc. London, vol. A247, pp. 529--551, April 1955.
% \bibitem{b2} J. Clerk Maxwell, A Treatise on Electricity and Magnetism, 3rd ed., vol. 2. Oxford: Clarendon, 1892, pp.68--73.
% \bibitem{b3} I. S. Jacobs and C. P. Bean, ``Fine particles, thin films and exchange anisotropy,'' in Magnetism, vol. III, G. T. Rado and H. Suhl, Eds. New York: Academic, 1963, pp. 271--350.
% \bibitem{b4} K. Elissa, ``Title of paper if known,'' unpublished.
% \bibitem{b5} R. Nicole, ``Title of paper with only first word capitalized,'' J. Name Stand. Abbrev., in press.
% \bibitem{b6} Y. Yorozu, M. Hirano, K. Oka, and Y. Tagawa, ``Electron spectroscopy studies on magneto-optical media and plastic substrate interface,'' IEEE Transl. J. Magn. Japan, vol. 2, pp. 740--741, August 1987 [Digests 9th Annual Conf. Magnetics Japan, p. 301, 1982].
% \bibitem{b7} M. Young, The Technical Writer's Handbook. Mill Valley, CA: University Science, 1989.
% \end{thebibliography}
% \vspace{12pt}
% \color{red}
% IEEE conference templates contain guidance text for composing and formatting conference papers. Please ensure that all template text is removed from your conference paper prior to submission to the conference. Failure to remove the template text from your paper may result in your paper not being published.

\begin{figure*}
\huge
\begin{center}
Supplementary Material for the paper "Optimal Update in\\ Energy Harvesting Aided Terahertz Communications with Random Blocking"    
\end{center}
\end{figure*}

\newpage

\section{APPENDIX}
\subsection{Proof of Lemma \ref{lemma1}}
\label{app_proof_lemma_monitonic}
    The proof requires the use of value iteration algorithm(VIA) and mathematical induction. First, give a brief introduction to VIA, which obtains the value of the value function in different states through continuous iteration. The specific iteration process is as follows:
\begin{equation}
\label{VIA}
\begin{cases}
{{V_0}(\mathbf{x}) = {m_\mathbf{x}}},\\
    Q_k(\mathbf{x},a)=C(\mathbf{x},a) +\underset{\mathbf{x}^\prime \in \mathcal{S}}{\sum}   \Pr (\mathbf{x}^\prime|\mathbf{x},a)V_k(\mathbf{x}^\prime),\\
{{V_{k + 1}}(\mathbf{x}) = \mathop {\min }\limits_{a \in \mathcal{A}} Q_k(\mathbf{x},a)},
\end{cases}
\end{equation}
where $m_\mathbf{x}$ is an arbitrary initial value of $V_0(\mathbf{x})$ with respect to state $\mathbf{x}$ and $k \in \mathbb Z$. It’s worth noting that $V_{k + 1}(\mathbf{x})$ will converge when $k$ goes into infinity for any state $\mathbf{x}$, which can be expressed as follows:
\begin{equation}
\label{Vkinfintiy}
\mathop {\lim }\limits_{k \to \infty } {V_k}(\mathbf{x}) = V(\mathbf{x}),\forall \mathbf{x} \in \mathcal{S}.
\end{equation}

Then we will use mathematical induction to prove the monotonicity of the value function in each component.  

First prove \eqref{lemma1_part1}. At the beginning of the induction method, We need to verify that the inequality $V_1(\Delta_1,q)\le V_1(\Delta_2,q)$  holds when $k=1$. By assuming ${{V_0}(\mathbf{x}) = 0},\forall \mathbf{x} \in \mathcal{S}$, we have:
\begin{align}
\label{V_0_DELTA1_lemma1}
    V_1(\Delta_1,q) &= \min_{a \in \mathcal{A}}\left\{ Q_0(\Delta_1,q,a)\right\} \nonumber\\
    &=\min\left\{ Q_0(\Delta_1,q,0),Q_0(\Delta_1,q,1)\right\} \nonumber\\
    &=\min\left\{ \Delta_1+\mathbb \omega{C_r} \mathds{1}(q),\Delta_1\right\} \nonumber\\
    &=\Delta_1,
\end{align}
and,
\begin{align}
\label{V_0_DELTA1_lemma2}
    V_1(\Delta_2,q) &= \min_{a \in \mathcal{A}}\left\{ Q_0(\Delta_2,q,a)\right\} \nonumber\\
    &=\min\left\{ Q_0(\Delta_2,q,0),Q_0(\Delta_2,q,1)\right\} \nonumber\\
    &=\min\left\{ \Delta_2+\omega{C_r} \mathds{1}(q),\Delta_2\right\} \nonumber\\
    &=\Delta_2.
\end{align}

Therefore, if $\Delta_1\le\Delta_2$, $V_1(\Delta_1,q)=\Delta_1\le\Delta_2=V_1(\Delta_2,q)$.
Then we assume that at the $k$th step of the induction method, the following formula holds:
\begin{equation}
\label{klimit}
    V_k(\Delta_1,q)\le V_k(\Delta_2,q),\forall \Delta_1\le\Delta_2.
\end{equation}
So the next formula that needs to be verified is
\begin{equation}
\label{V_k+1}
    V_{k+1}(\Delta_1,q)\le V_{k+1}(\Delta_2,q),\forall \Delta_1\le\Delta_2
\end{equation}
Since ${{V_{k + 1}}(\mathbf{x}) = \mathop {\min }\limits_{a \in \mathcal{A}} Q_k(\mathbf{x},a)}$, we need to bring out $Q_k(\mathbf{x},a)$ first. The state-action value function $Q_k(\mathbf{x},a)$ at state $\mathbf{x}=(\Delta, q)$ is as follows:
\begin{equation}
\label{Q_k_equation}
\begin{cases}
 Q_k(\Delta, q,0)=C(\Delta, q,0) +\underset{\mathbf{x}^\prime \in \mathcal{S}}{\sum}   \Pr (\mathbf{x}^\prime|\mathbf{x},0)V_k(\mathbf{x}^\prime),\\
    Q_k(\Delta, q,1)=C(\Delta, q,1) +\underset{\mathbf{x}^\prime \in \mathcal{S}}{\sum}   \Pr (\mathbf{x}^\prime|\mathbf{x},1)V_k(\mathbf{x}^\prime).
\end{cases}
\end{equation}
Due to the complexity of the transition probability situation and one-step cost function, we will discuss the following three cases:

\textbf{\emph{Case 1}}. $q=0$,

In this case, according to transition probability \eqref{transition_case1_v2} and \eqref{transition_case2_v2}, we have the state-value function $Q_k(\Delta, q,0)$ and $Q_k(\Delta, q,1)$ as follows:
\begin{align}
    Q_k(\Delta, q,0)=\Delta &+\lambda V_k(\Delta+1,q+1) \nonumber\\
    &+(1- \lambda)V_k(\Delta+1,q),
\end{align}
and,
\begin{align}
    Q_k(\Delta, q,1)=\Delta+\omega C_r &+p\lambda V_k(\Delta+1,q+1) \nonumber\\
    &+p(1- \lambda)V_k(\Delta+1,q)\nonumber\\
    &+(1-p)\lambda V_k(1,q+1)\nonumber\\
    &+(1-p)(1-\lambda) V_k(1,q).
\end{align}
Due to that $V_{k}(\Delta,q)$ is assumed to be non-decreasing function with respect to $\Delta$ for any fixed $q$, it is obviously that both $Q_k(\Delta, q,0)$ and $Q_k(\Delta, q,1)$ are non-decreasing with respect to $\Delta$. Therefore,for any $\Delta_1\le\Delta_2$ we have:
\begin{align}
\label{V_K+1_DELTA1_lemma1}
    V_{k+1}(\Delta_1,q) &= \min_{a \in \mathcal{A}}\left\{ Q_k(\Delta_1,q,a)\right\} \nonumber\\
    &=\min\left\{ Q_k(\Delta_1,q,0),Q_k(\Delta_1,q,1)\right\} \nonumber\\
    &\le \min\left\{ Q_k(\Delta_2,q,0),Q_k(\Delta_2,q,1)\right\} \nonumber\\
    &=V_{k+1}(\Delta_2,q).
\end{align}
%So \eqref{V_k+1} still holds. 
As a result, with the induction we prove that $V_{k}(\Delta,q)$ is non-decreasing function for any $k$ with respect to $\Delta$ and $q=0$. By taking the limits on both side of \eqref{klimit} we prove that \eqref{lemma1_part1} holds in the case $q=0$.

\textbf{\emph{Case 2}}. $0<q<B$,

In this case, according to transition probability \eqref{transition_case1_v2} and \eqref{transition_case2_v2}, we have the state-value function $Q_k(\Delta, q,0)$ and $Q_k(\Delta, q,1)$ as follows:
\begin{align}
    Q_k(\Delta, q,0)=\Delta &+\lambda V_k(\Delta+1,q+1) \nonumber\\
    &+(1- \lambda)V_k(\Delta+1,q),
\end{align}
and,
\begin{align}
    Q_k(\Delta, q,1)=\Delta &+p\lambda V_k(\Delta+1,q) \nonumber\\
    &+p(1- \lambda)V_k(\Delta+1,q-1)\nonumber\\
    &+(1-p)\lambda V_k(1,q)\nonumber\\
    &+(1-p)(1-\lambda) V_k(1,q-1).
\end{align}
Due to $V_{k}(\Delta,q)$ is assumed to be non-decreasing function with respect to $\Delta$ for any fixed $q$, it is obviously that both $Q_k(\Delta, q,0)$ and $Q_k(\Delta, q,1)$ are non-decreasing with respect to $\Delta$. Therefore,for any $\Delta_1\le\Delta_2$ we have:
\begin{align}
\label{V_K+1_DELTA1_lemma1_case2}
    V_{k+1}(\Delta_1,q) &= \min_{a \in \mathcal{A}}\left\{ Q_k(\Delta_1,q,a)\right\} \nonumber\\
    &=\min\left\{ Q_k(\Delta_1,q,0),Q_k(\Delta_1,q,1)\right\} \nonumber\\
    &\le \min\left\{ Q_k(\Delta_2,q,0),Q_k(\Delta_2,q,1)\right\} \nonumber\\
    &=V_{k+1}(\Delta_2,q).
\end{align}
%So \eqref{V_k+1} still holds. 
As a result, with the induction we prove that $V_{k}(\Delta,q)$ is non-decreasing function for any $k$ with respect to $\Delta$ and any $q \in \left\{1,...,B-1\right\}$. By taking the limits on both side of \eqref{klimit} we prove that \eqref{lemma1_part1} holds in the case $0<q<B$.

\textbf{\emph{Case 3}}. $q=B$,

In this case, according to transition probability \eqref{transition_case1_v2} and \eqref{transition_case2_v2}, we have the state-value function $Q_k(\Delta, q,0)$ and $Q_k(\Delta, q,1)$ as follows:
\begin{align}
    Q_k(\Delta, q,0)=\Delta &+\lambda V_k(\Delta+1,q) \nonumber\\
    &+(1- \lambda)V_k(\Delta+1,q),
\end{align}
and,
\begin{align}
    Q_k(\Delta, q,1)=\Delta+\omega C_r &+p\lambda V_k(\Delta+1,q) \nonumber\\
    &+p(1- \lambda)V_k(\Delta+1,q-1)\nonumber\\
    &+(1-p)\lambda V_k(1,q)\nonumber\\
    &+(1-p)(1-\lambda) V_k(1,q-1).
\end{align}
Due to $V_{k}(\Delta,q)$ is assumed to be non-decreasing function with respect to $\Delta$ for any fixed $q$, it is obviously that both $Q_k(\Delta, q,0)$ and $Q_k(\Delta, q,1)$ are non-decreasing with respect to $\Delta$. Therefore,for any $\Delta_1\le\Delta_2$ we have:
\begin{align}
\label{V_K+1_DELTA1_lemma1_case3}
    V_{k+1}(\Delta_1,q) &= \min_{a \in \mathcal{A}}\left\{ Q_k(\Delta_1,q,a)\right\} \nonumber\\
    &=\min\left\{ Q_k(\Delta_1,q,0),Q_k(\Delta_1,q,1)\right\} \nonumber\\
    &\le \min\left\{ Q_k(\Delta_2,q,0),Q_k(\Delta_2,q,1)\right\} \nonumber\\
    &=V_{k+1}(\Delta_2,q).
\end{align}
%So \eqref{V_k+1} still holds. 
As a result, with the induction we prove that $V_{k}(\Delta,q)$ is non-decreasing function for any $k$ with respect to $\Delta$ and $q=B$. By taking the limits on both side of \eqref{klimit} we prove that \eqref{lemma1_part1} holds in the case $q=B$.

To sum up, \eqref{lemma1_part1} holds and we complete the proof of the first part in Lemma\ref{lemma1}.

According to the exact same mathematical induction, we can also verify that the formula \eqref{lemma1_part2} holds. Due to limited space, the specific certification steps are omitted here. Thus we complete the proof of Lemma \ref{lemma1}.

\subsection{Proof of Lemma \ref{lemma2}}
\label{app_proof_lemma_creasement}

First, let's prove \eqref{lemm2_formula1}. By the \eqref{lemma1_part1} of Lemma \ref{lemma1}, assuming $\Delta_1 \le \Delta_2$ and $q \in \left\{1,...,B-1\right\}$, it is easy to yield
    \begin{align}
        Q(\Delta_2,q,0)&-Q(\Delta_1,q,0) =\Delta_2 - \Delta_1 \nonumber\\
        &+\lambda[V(\Delta_2+1,q+1)-V(\Delta_1+1,q+1)]\nonumber\\
        &+(1-\lambda)[V(\Delta_2+1,q)-V(\Delta_1+1,q)]\nonumber\\
        \ge &\Delta_2 - \Delta_1,
    \end{align}
    and,
    \begin{align}
        Q(\Delta_2,&q,1)-Q(\Delta_1,q,1) =\Delta_2 - \Delta_1 \nonumber\\
        &+p\lambda[V(\Delta_2+1,q)-V(\Delta_1+1,q)]\nonumber\\
        &+p(1-\lambda)[V(\Delta_2+1,q-1)-V(\Delta_1+1,q-1)]\nonumber\\
        &+(1-p)\lambda[V(1,q)-V(1,q)]\nonumber\\        
        &+(1-p)(1-\lambda)[V(1,q-1)-V(1,q-1)]\nonumber\\
        \ge &\Delta_2 - \Delta_1.
    \end{align}
Due to $V(\mathbf{x})= \mathop{\min}\limits_{a \in \mathcal{A}} Q(\mathbf{x},a)$, we prove that formula \eqref{lemm2_formula1} holds for all $q \in \left\{1,...,B-1\right\}$. Through the same proof process, it can also be verified that \eqref{lemm2_formula1} is also valid when $q=0$ and $q=B$. Therefore, we have proved $V(\Delta_2,q)-V(\Delta_1,q)\ge \Delta_2-\Delta_1$ holds for any $\Delta_1 \le \Delta_2$ and fixed $q \in \mathcal{B}$.

Second, we will tackle formula \eqref{lemm2_formula2}. The following proof needs to apply VIA and mathematical induction. For the convenience of explanation, an equivalent transformation is made to formula \eqref{lemm2_formula2} as follows:
    \begin{equation}
    \label{trans_lemma2_formula2}
        V(\Delta+1,q+1)+p V(\Delta,q)\ge V(\Delta,q+1)+p V(\Delta+1,q),
    \end{equation}
    
for state $\mathbf{x}$, we have 
\begin{align}
V(\mathbf{x})&= \mathop{\min}\limits_{a \in \mathcal{A}} Q(\mathbf{x},a)\nonumber\\
&=\min\left\{ Q_k(\mathbf{x},0),Q_k(\mathbf{x},1)\right\}.
\end{align}

So every value function in \eqref{trans_lemma2_formula2} has two possible values. In order to prove formula \eqref{trans_lemma2_formula2}, theoretically we need to discuss $2^4=16$ cases, which is obviously a bit too cumbersome. Here we use a little trick, that is, as long as we prove that for the $2^2=4$ possible combinations on the left side of the inequality sign, there exists a combination on the right side of the inequality sign to make "$\ge$" hold, then we can prove formula \eqref{trans_lemma2_formula2}. Next, we make a mapping, using four numbers to sequentially represent the action taken by the minimum state-action value function in formula \eqref{trans_lemma2_formula2}, that is, "1010" represents the following:
    \begin{align}
    \label{1010}
        Q(\Delta+1,q+1,1)+pQ(\Delta,q,0)\ge \nonumber\\ Q(\Delta,q+1,1)+pQ(\Delta+1,q,0),
    \end{align}
So according to the previous trick, we only need to verify "0000", "1010", "0101", "1111" to prove formula \eqref{trans_lemma2_formula2}. Due to limited space, we only show the verification process of "1010" in the following proof. The other three cases can also be proved by the same steps.

Now we start to apply VIA. Assuming that $V_0(\mathbf{x})=0$ for any states $\mathbf{x}$, we have:
\begin{align}
    &Q_0(\Delta+1,q+1,1)+p Q_0(\Delta,q,0)\nonumber\\
    &-[Q_0(\Delta,q,1)+p Q_0(\Delta+1,q,0)]\nonumber\\
    =&\Delta+1+p(\Delta+\omega\mathds{1} (q)C_r)-[\Delta+p(\Delta+\omega\mathds{1} (q)C_r)]\nonumber\\
    =&1 \ge 0.
\end{align}
Then for $Q_0(\mathbf{x})$ we can also verify the same property in the "0000", "0101", "1111" case by the similar calculation, which implies:
\begin{equation}
\label{V_1_trans_lemma2_formula2}
    V_1(\Delta+1,q+1)+p V_1(\Delta,q)\ge V_1(\Delta,q+1)+p V_1(\Delta+1,q),
\end{equation}
for any $q \in \left\{0,1,...,B-1\right\}$ and $\Delta \in \mathbb Z^+ $. By induction, assuming that for any $q \in \left\{0,1,...,B-1\right\}$ and $\Delta \in \mathbb Z^+ $, we have:
\begin{equation}
\label{V_k_trans_lemma2_formula2}
    V_k(\Delta+1,q+1)+p V_k(\Delta,q)\ge V_k(\Delta,q+1)+p V_k(\Delta+1,q).
\end{equation}
What we need to do is to verify that formula \eqref{trans_lemma2_formula2} still holds in the next value iteration. Again, we take a look at the "1010" case. For $\Delta \in \mathbb Z^+ $ and $q \in \left\{0,1,...,B-1\right\}$, we have:
% since the transition probability and one-step cost is related to the value of $q$, we need to discuss it in the following two sub-cases:
    \begin{align}
    \label{Qk11010}
        &Q_k(\Delta+1,q+1,1)+pQ_k(\Delta,q,0)\nonumber\\
        &-[Q_k(\Delta,q+1,1)+pQ_k(\Delta+1,q,0)]\nonumber\\
        =&\Delta+1+p\lambda V_k(\Delta+2,q+1)+p(1- \lambda)V_k(\Delta+2,q)\nonumber\\
    &+(1-p)\lambda V_k(1,q+1)+(1-p)(1-\lambda) V_k(1,q)\nonumber\\
    &+p[\Delta+\omega C_r+\lambda V_k(\Delta+1,q+1)+(1- \lambda)V_k(\Delta+1,q)]\nonumber\\
    &-\Delta-p\lambda V_k(\Delta+1,q+1)-p(1- \lambda)V_k(\Delta+1,q)\nonumber\\
    &-(1-p)\lambda V_k(1,q+1)-(1-p)(1-\lambda) V_k(1,q)\nonumber\\
    &-p[\Delta+1+\omega C_r+\lambda V_k(\Delta+2,q+1)-(1- \lambda)V_k(\Delta+2,q)]\nonumber\\
    =&1-p\ge 0.
    \end{align}

%Here we first check the case $q \in \left\{1,...,B-2\right\}$. When $k=1$, we have:
%    \begin{align}
    % \label{Q1010}
    %     &Q(\Delta+1,q+1,1)+pQ(\Delta,q,0)\nonumber\\
    %     &-[Q(\Delta,q+1,1)+pQ(\Delta+1,q,0)]\nonumber\\
    %     =&\Delta+1+p\lambda V(\Delta+2,q+1)+p(1- \lambda)V(\Delta+2,q)\nonumber\\
    % &+(1-p)\lambda V(1,q+1)+(1-p)(1-\lambda) V(1,q)\nonumber\\
    % &+p[\Delta+\lambda V(\Delta+1,q+1)+(1- \lambda)V(\Delta+1,q)]\nonumber\\
    % &-\Delta-p\lambda V(\Delta+1,q+1)-p(1- \lambda)V(\Delta+1,q)\nonumber\\
    % &-(1-p)\lambda V(1,q+1)-(1-p)(1-\lambda) V(1,q)\nonumber\\
    % &-p[1+\lambda V(\Delta+2,q+1)-(1- \lambda)V(\Delta+2,q)]\nonumber\\
    % =&1-p\ge 0
    % \end{align}
Therefore, by the similar step, we can verify the other three cases and get the following formula 
\begin{equation}
\label{V_k+1_trans_lemma2_formula2}
    V_{k+1}(\Delta+1,q+1)+p V_{k+1}(\Delta,q)\ge V_{k+1}(\Delta,q+1)+p V_{k+1}(\Delta+1,q)
\end{equation}
holds for any $\Delta \in \mathbb Z^+ $ and $q \in \left\{0,1,...,B-1\right\}$. By induction we confirm that for any $k$, the formula  \eqref{V_k_trans_lemma2_formula2} holds. Take the limits of $k$ on both side then we are able to prove that \eqref{trans_lemma2_formula2} holds, which is equivalent to \eqref{lemm2_formula2} holds. Hence, we complete the whole proof.
% $\Delta=0$, formula \eqref{trans_lemma2_formula2} holds for any $q \in \left\{1,...,B-2\right\}$. Thus by induction, we assume that formula \eqref{trans_lemma2_formula2} holds for $\Delta=k$ and $q \in \left\{1,...,B-2\right\}$,where $k$ are positive integers greater than 1. The next job is to prove that when $\Delta=k+1$, the formula \eqref{trans_lemma2_formula2} still holds. Here we also check the case "1010":
    % \begin{align}
    % \label{K+11010}
    %     Q(k&+2,q+1,1)+pQ(k+1,q,0)\nonumber\\
    %     &-[Q(k+1,q+1,1)+pQ(k+2,q,0)]\nonumber\\
    %     &=k+2+p\lambda V(k+3,q+1)+p(1- \lambda)V(k+3,q)\nonumber\\
    % &+(1-p)\lambda V(1,q+1)+(1-p)(1-\lambda) V(1,q)\nonumber\\
    % &+p[k+1+\lambda V(k+2,q+1)+(1- \lambda)V(k+2,q)]\nonumber\\
    % &-k-1-p\lambda V(k+2,q+1)-p(1- \lambda)V(k+2,q)\nonumber\\
    % &-(1-p)\lambda V(1,q+1)-(1-p)(1-\lambda) V(1,q)\nonumber\\
    % &-p[k+2+\lambda V(k+3,q+1)-(1- \lambda)V(k+3,q)]\nonumber\\
    % &=1-p\ge 0
    % \end{align}

\end{document}